\documentclass{aa}
\usepackage{amsbsy}
\usepackage{amssymb}
\usepackage[dvips]{graphicx}

\usepackage{txfonts}
\usepackage{natbib}

\begin{document}

\title{
  Constraining the evolution of dark energy with type Ia supernovae
  and gamma-ray bursts
}

\author{
  Shi Qi\inst{1,4}
  \and
  Fa-Yin Wang\inst{2}
  \and
  Tan Lu\inst{3,4}
}
\institute{
  Department of Physics, Nanjing University, Nanjing 210093, China \\
  \email{qishi11@gmail.com}
  \and
  Department of Astronomy, Nanjing University, Nanjing 210093, China \\
  \email{fayinwang@nju.edu.cn}
  \and
  Purple Mountain Observatory, Chinese Academy of Sciences, Nanjing
  210008, China \\
  \email{t.lu@pmo.ac.cn}
  \and
  Joint Center for Particle, Nuclear Physics and Cosmology, Nanjing
  University -- Purple Mountain Observatory, Nanjing  210093, China
}

\abstract{}{
  The behavior of the dark energy equation of state (EOS) is
  crucial in distinguishing different cosmological
  models. With a model independent approach, we constrain the
  possible evolution of the dark energy EOS.
}{
  Gamma-ray bursts (GRBs) of redshifts up to $z>6$ are used, in
  addition to type Ia
  supernovae (SNe Ia). We separate the redshifts into $4$ bins
  and assume a constant EOS parameter for dark energy in each bin.
  The EOS parameters are decorrelated by diagonalizing the covariance
  matrix. And the evolution of dark energy is estimated out of the
  uncorrelated EOS parameters.
}{
  By including GRB luminosity data, we significantly reduce the
  confidence
  interval of the uncorrelated EOS parameter whose contribution
  mostly comes from the redshift bin of $0.5<z<1.8$. At high redshift
  where we only have GRBs, the constraints on the dark energy EOS
  are still very weak. However, we can
  see an obvious cut at about zero in the probability plot of
  the EOS parameter, from which we can infer that the ratio of dark
  energy to matter most probably continues to decrease beyond redshift
  $1.8$. We carried out analyses with and without including the latest
  BAO measurements, which themselves favor a dark energy EOS of
  $w<-1$. If they are included, the results show some evidence of an
  evolving dark energy EOS. If not included, however, the
  results are consistent with
  the cosmological constant within $1 \sigma$ for redshift $0<z
  \lesssim 0.5$ and $2 \sigma$ for $0.5 \lesssim z<1.8$.
}{}

\keywords{cosmological parameters - supernovae: general - Gamma rays:
  bursts}

\maketitle

\section{Introduction}

Unexpected accelerating expansion of the universe
was first discovered by observing type Ia supernovae (SNe
Ia)~\citep{Riess:1998cb,Perlmutter:1998np}.
This acceleration is attributed to dark energy, whose presence was
corroborated later by other independent sources including
the WMAP and other observations of the CMB~\citep{Spergel:2003cb},
X-ray clusters~\citep{Allen:2002sr}, etc.
With more observational data
available~\citep[e.g.][]{Hawkins:2002sg, Abazajian:2003jy,
  Spergel:2006hy, Riess:2006fw, WoodVasey:2007jb, Davis:2007na,
  Schaefer:2006pa, Percival:2007yw, Komatsu:2008hk, Dunkley:2008ie},
we are getting more stringent constraints on the nature of dark
energy; nevertheless, the underlying physics of dark energy remains
mysterious. In addition to the cosmological constant, many other dark
energy models have been suggested, including models of scalar fields
(see~\citet{Copeland:2006wr} for a recent review) and
modification of general
relativity~\citep[see for example][]{Deffayet:2000uy, Binetruy:1999hy,
  Maartens:2006qf, Capozziello:2003gx, Dvali:2000rv, Carroll:2003wy,
  Nojiri:2003ft, Nojiri:2006ri}.

Measuring the expansion history directly may be the best way to
constrain the properties of dark energy. To measure the expansion
history, we need standard candles at different redshifts. SNe Ia,
which are now viewed as nearly ideal standard candles, have played an
important role in 
constraining cosmological parameters. We now have $192$ samples of SN 
Ia~\citep{Riess:2006fw,WoodVasey:2007jb,Davis:2007na} that can be used
to determine the expansion history. And the proposed SNAP
satellite\footnote{See http://snap.lbl.gov/}~\citep[see for
example][]{Aldering:2004ak}
will add about $2000$ samples per year. 
Increasing SN Ia samples will provide more and more precise
description of the cosmic expansion. However, the redshift of the
present $192$ SNe Ia ranges only up to about $1.7$ and the mean
redshift is about $0.5$. They cannot provide any information on the
cosmic expansion beyond redshift $1.7$.
Here gamma-ray bursts (GRBs) come in and fill the void. With their
higher luminosities, GRBs are visible across much greater distances
than supernovae. The presently available $69$ compiled
GRBs~\citep{Schaefer:2006pa} extend the redshift to $z>6$ and the mean
redshift is about $2.1$. After being calibrated
with luminosity relations, GRBs may be
used as standard candles to provide information on cosmic expansion at
high redshift and, at the same time, to tighten the constraints on
cosmic expansion at low redshift.
See, for example,~\citet{Dai:2004tq}, \citet{Ghirlanda:2004fs},
\citet{DiGirolamo:2005ze}, \citet{Firmani:2005gs},
\citet{Friedman:2004mg}, \citet{Lamb:2005cw},
\citet{Liang:2005xb}, \citet{Xu:2005uv}, \citet{Wang:2005ic},
\citet{Li:2006ev}, \citet{Su:2006jp}, \citet{Schaefer:2006pa},
\citet{Wright:2007vr}, and \citet{Wang:2007rz}
for works on GRB cosmology.

Among parameters that describe the properties of dark energy, the
equation of state (EOS) is the most important.
Whether and how it evolves with time is crucial in distinguishing
different cosmological models. Due to not
understanding of the behaviors of dark energy, simple parametric
forms such as
$w(z)=w_0+w'z$~\citep{Cooray:1999da} and
$w(z)=w_0+w_az/(1+z)$~\citep{Chevallier:2000qy, Linder:2002et}
have been proposed for studying the possible evolution of dark
energy. However, a simple parameterization itself
greatly restricts the allowed wandering of $w(z)$, and is equivalent
to a strong prior on the nature of dark energy~\citep{Riess:2006fw}.
To avoid any strong prior before comparing data, one can utilize
an alternative approach in which uncorrelated estimates are made of
discrete $w(z)$ of different redshifts. This approach was proposed
by~\citet{Huterer:2002hy} and \citet{Huterer:2004ch} and has been
adopted in previous analyses using SNe
Ia~\citep{Riess:2006fw,Sullivan:2007pd}.

In this work, we apply this approach to GRB luminosity
data~\citep{Schaefer:2006pa}, in addition to SN Ia
data~\citep{Riess:2006fw,WoodVasey:2007jb,Davis:2007na}, and compare
our results with those in the previous work that does not include
GRB luminosity data~\citep{Sullivan:2007pd}.
We first briefly review the techniques for uncorrelated estimates of
dark energy evolution in section~\ref{sec:methodology}.
The observational data and how they are included in the data analysis
are described in section~\ref{sec:observational_data}.
We present our results in section~\ref{sec:results}, followed by a
summary in section~\ref{sec:summary}.

\section{Methodology}
\label{sec:methodology}

Standard candles impose constraints on cosmological parameters
essentially through a comparison of the luminosity distance from
observation with that from theoretical models. Observationally, the
luminosity distance is given by
\begin{equation}
  \label{eq:dlz_observation}
  d_L=\left(
    \frac{L}{4\pi F}
  \right)^{1/2}
  ,
\end{equation}
where $L$ and $F$ are the luminosity of the standard candles and the
observed flux, respectively.
Theoretically, the luminosity distance $d_L(z)$ depends on
the geometry of the universe, i.e. the sign of $\Omega_k$, and is
given by
\begin{equation}
  \label{eq:dlz_theoretical}
  d_L(z)=(1+z)\frac{c}{H_0}\times
  \left\{
    \begin{array}{ll}
      \frac{1}{\sqrt{|\Omega_k|}}
      \sinh\left(
        \sqrt{|\Omega_k|} \int_0^z\frac{\mathrm{d}\tilde{z}}{E(\tilde{z})}
      \right) & \textrm{if } \Omega_k>0
      \\
      \int_0^z\frac{\mathrm{d}\tilde{z}}{E(\tilde{z})} 
      & \textrm{if } \Omega_k=0
      \\
      \frac{1}{\sqrt{|\Omega_k|}}
      \sin\left(
        \sqrt{|\Omega_k|} \int_0^z\frac{\mathrm{d}\tilde{z}}{E(\tilde{z})}
      \right) & \textrm{if } \Omega_k<0
    \end{array}
  \right.
  ,
\end{equation}
where
\begin{eqnarray}
  \label{eq:Ez}
  E(z)=\left[
    \Omega_m (1+z)^3+\Omega_x f(z) + \Omega_k (1+z)^2
  \right]^{1/2},
  &&
  \nonumber \\
  \Omega_m+\Omega_x+\Omega_k=1
  &&
\end{eqnarray}
and
\begin{eqnarray}
  \label{eq:fz}
  f(z)=\exp \left[
    3\int_0^z\frac{1+w(\tilde{z})}{1+\tilde{z}}\mathrm{d}\tilde{z}
  \right]
  .
\end{eqnarray}
Dark energy parameterization schemes enter through $f(z)$. For the
case where EOS is piecewise constant in redshift, $f(z)$ can be
rewritten as~\citep{Sullivan:2007pd}
\begin{equation}
  \label{eq:fzbinned}
  f(z_{n-1}<z \le z_n)=
  (1+z)^{3(1+w_n)}\prod_{i=0}^{n-1}(1+z_i)^{3(w_i-w_{i+1})}
  ,
\end{equation}
where $w_i$ is the EOS parameter in the $i^{\mathrm{th}}$ redshift bin
defined by
an upper boundary at $z_i$, and the zeroth bin is defined as $z_0=0$.
In order to compare with previous analysis~\citep{Sullivan:2007pd}, we
define the first three redshift bins to be the same as those
used by~\citet{Sullivan:2007pd} by setting $z_1=0.2$, $z_2=0.5$, and
$z_3=1.8$. The fourth bin is defined by $z_4=7$ to include GRBs.
We carry out our analyses under two different assumptions about the
high redshift (redshift greater than $z_4=7$ in our case)
behavior of dark energy, i.e. the so-called~\citep[see][]{Riess:2006fw}
``weak'' prior, which makes no assumptions about $w(z)$ at $z>7$ and
the ``strong'' prior, which assumes $w(z)=-1$ at $z>7$.

In this paper we adopt $\chi^2$ statistic to estimate parameters. For
a physical quantity $\xi$ with experimentally measured value $\xi_o$,
standard deviation $\sigma_{\xi}$, and theoretically predicted value
$\xi_t(\theta)$, where $\theta$ is a collection of parameters needed
to calculate the theoretical value, the $\chi^2$ value is given by
\begin{equation}
  \label{eq:chi2_xi}
  \chi_{\xi}^2(\theta)=\frac{
    \left(
      \xi_t(\theta)-\xi_o
    \right)^2
  }{\sigma_{\xi}^2}
\end{equation}
and the total $\chi^2$ is the sum of all $\chi_{\xi}^2$s,
i.e.
\begin{equation}
  \label{eq:chi2}
  \chi^2(\theta)=\sum_{\xi}\chi_{\xi}^2(\theta)
  .
\end{equation}
The likelihood function is then proportional to
$\exp\left(-\chi^2(\theta)/2\right)$, which produces the posterior
probability when multiplied by the prior probability of $\theta$. In
the case of our analysis, the calculation of
$\chi^2$s for different observational data is described in
section~\ref{sec:observational_data}. According to the posterior
probability derived in this way, Markov chains are generated through
the Monte-Carlo algorithm to study the statistical properties of the
parameters. In this paper, we focus on the EOS parameters by
marginalizing the others.

As mentioned above, in the process of constraining cosmological
parameters, standard candles play this role by providing the
luminosity distances at certain
redshifts. However, the luminosity distance depends on the integration
of the behavior of the dark energy over redshift, so the estimates of
the dark energy EOS
parameters $w_i$ at high redshift depend on those at
low redshift. In other words, the EOS parameters $w_i$ are correlated
in the sense that the covariance matrix,
\begin{equation}
  \label{eq:cov_matrix}
  \boldsymbol{C}
  =\langle \boldsymbol{w} \boldsymbol{w}^{\mathrm{T}} \rangle 
  - \langle \boldsymbol{w} \rangle 
  \langle \boldsymbol{w}^{\mathrm{T}} \rangle
  ,
\end{equation}
is not diagonal. In the above equation, the $\boldsymbol{w}$ is a
vector with components $w_i$ and the average is calculated by letting
$\boldsymbol{w}$ run over the Markov chain.
We can obtain a set of decorrelated parameters $\widetilde{w}_i$
through diagonalization of the covariance matrix by choosing an
appropriate transformation
\begin{equation}
  \label{eq:transformation}
  \widetilde{\boldsymbol{w}}=\boldsymbol{T} \boldsymbol{w}
  .
\end{equation}
There can be different choices for $\boldsymbol{T}$. In this paper we
use the transformation advocated by~\citet{Huterer:2004ch} (see below).
First we define the Fisher matrix
\begin{equation}
  \label{eq:fisher_matrix}
  \boldsymbol{F}\equiv\boldsymbol{C}^{-1}
  =\boldsymbol{O}^{\mathrm{T}}\boldsymbol{\Lambda}\boldsymbol{O}
  ,
\end{equation}
and then the transformation matrix $\boldsymbol{T}$ is given by
\begin{equation}
  \label{eq:transf_matrix1}
  \boldsymbol{T}=\boldsymbol{O}^{\mathrm{T}}
  \boldsymbol{\Lambda}^{\frac{1}{2}}\boldsymbol{O}
  ,
\end{equation}
except that the rows of the matrix $\boldsymbol{T}$ are normalized
such that
\begin{equation}
  \label{eq:transf_matrix2}
  \sum_j T_{ij}=1
  .
\end{equation}
The advantage of this transformation is that the weights (rows of
$\boldsymbol{T}$) are positive almost everywhere and localized in
redshift fairly well, so the uncorrelated EOS parameters
$\widetilde{w}_i$ are easy to interpret
intuitively~\citep{Huterer:2004ch}.

\section{Observational data}
\label{sec:observational_data}

To constrain the dark energy EOS, we have made use of observational
data described below.

\subsection{Type Ia supernovae}

\begin{figure}[htbp]
  \centering
  \includegraphics[width=0.5\textwidth]{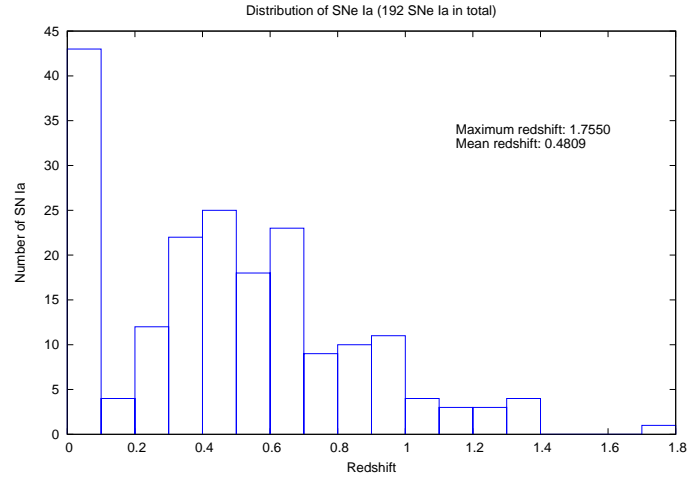}
  \caption{Distribution of SN Ia samples versus redshift}
  \label{fig:SNIa_distr}
\end{figure}

Recently compiled SN Ia
data~\citep{Riess:2006fw,WoodVasey:2007jb,Davis:2007na} include 45
nearby supernovae~\citep{Hamuy:1996su, Riess:1998dv, Jha:2005jg}, 60
ESSENCE supernovae~\citep{WoodVasey:2007jb}, 57 SNLS
supernovae~\citep{Astier:2005qq}, and 30 HST
supernovae~\citep{Riess:2006fw}. Figure~\ref{fig:SNIa_distr} shows the
distribution of these SN Ia samples versus redshift.
The $\chi^2$ value for SNe Ia is
\begin{equation}
  \label{eq:chi2_SNIa}
  \chi^2_{\mathrm{SN}}=\sum_i \frac{
    (\mu_{p,i}-\mu_{o,i})^2
  }{
    \sigma_i^2+\sigma_{int}^2
  }
  ,
\end{equation}
where $\mu_{o,i}$ and $\mu_{p,i}$ are the observed and theoretically
predicted distance modulus of SN Ia, which is defined by
$\mu=5\log d_L+25$ with the luminosity distance $d_L$ in unit of
megaparsec and $\sigma_{int}$ is the intrinsic dispersion.

\subsection{Gamma-ray bursts}

Besides SNe Ia, GRB luminosity data is another main observational
constraint we
used. As mentioned before, GRBs are complementary to SNe Ia at high
redshifts.
\begin{figure}[htbp]
  \centering
  \includegraphics[width=0.5\textwidth]{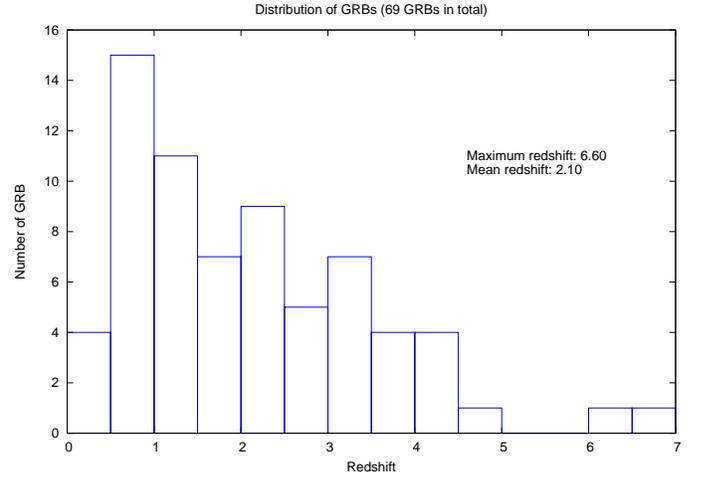}
  \caption{Distribution of GRB samples versus redshift}
  \label{fig:GRB_distr}
\end{figure}
We include GRBs presented by~\citet{Schaefer:2006pa} (see
Figure~\ref{fig:GRB_distr} for the distribution of these GRBs versus
redshift) in our
analysis by utilizing the five luminosity relations, i.e. the
connections between measurable parameters of the light curves and/or
spectra and GRB luminosity: $\tau_{\mathrm{lag}}$-$L$, $V$-$L$,
$E_{\mathrm{peak}}$-$L$, $E_{\mathrm{peak}}$-$E_{\mathrm{\gamma}}$ and
$\tau_{\mathrm{RT}}$-$L$
\begin{eqnarray}
  \label{eq:GRB-lag-L}
  \log \frac{L}{1 \; \mathrm{erg} \; \mathrm{s}^{-1}}
  &=& a_1+b_1 \log
  \left[
    \frac{\tau_{\mathrm{lag}}(1+z)^{-1}}{0.1\;\mathrm{s}}
  \right]
  ,
  \\
  \label{eq:GRB-V-L}
  \log \frac{L}{1 \; \mathrm{erg} \; \mathrm{s}^{-1}}
  &=& a_2+b_2 \log
  \left[
    \frac{V(1+z)}{0.02}
  \right]
  ,
  \\
  \label{eq:GRB-E_peak-L}
  \log \frac{L}{1 \; \mathrm{erg} \; \mathrm{s}^{-1}}
  &=& a_3+b_3 \log
  \left[
    \frac{E_{\mathrm{peak}}(1+z)}{300\;\mathrm{keV}}
  \right]
  ,
  \\
  \label{eq:GRB-E_peak-E_gamma}
  \log \frac{E_{\gamma}}{1\;\mathrm{erg}}
  &=& a_4+b_4 \log
  \left[
    \frac{E_{\mathrm{peak}}(1+z)}{300\;\mathrm{keV}}
  \right]
  ,
  \\
  \label{eq:GRB-tau_RT-L}
  \log \frac{L}{1 \; \mathrm{erg} \; \mathrm{s}^{-1}}
  &=& a_5+b_5 \log
  \left[
    \frac{\tau_{\mathrm{RT}}(1+z)^{-1}}{0.1\;\mathrm{s}}
  \right]
  .
\end{eqnarray}
Throughout this paper, by GRB luminosity data we refer to the GRBs'
observational data related to such luminosity relations. It is worth
mentioning that these relations may be correlated. As discussed
in~\citet{Schaefer:2006pa}, there is one significant correlation
between the $V$-$L$ and $\tau_{RT}$-$L$ relations with the correlation
coefficient equaling $0.53$. However, even for this correlation,
ignoring it only causes a 4\% underestimate in the standard error of
the average distance modulus~\citep{Schaefer:2006pa}, so in our
analysis we safely ignore the correlations and simply add the
contributions from each relation (see Eq.~(\ref{eq:chi2_GRB}) below).

There are significant differences between SNe Ia and GRBs on the
calibration. For SNe Ia, the calibration is done with nearby events
and is therefore independent of cosmological parameters.
The luminosity relations obtained in the calibration are applied to
high-redshift events to derive
the luminosity of SNe Ia, then used to constrain cosmological
parameters. In this procedure, the calibration and the constraining of
cosmological parameters are done separately.
In contrast to SNe Ia, to constrain cosmological parameters using
GRBs, we need to know the luminosity relations of
GRBs (Eq.~(\ref{eq:GRB-lag-L}-\ref{eq:GRB-tau_RT-L})), i.e. to know
the values of $a_1$-$a_5$ and $b_1$-$b_5$;
consequently, we need the luminosity $L$ and the total
collimation-corrected energy $E_{\gamma}$ of GRBs, which are
converted respectively from the bolometric peak flux
$P_{\mathrm{bolo}}$ and the bolometric fluence $S_{\mathrm{bolo}}$ of
GRBs through the relations
\begin{eqnarray}
  \label{eq:GRB-L-P_bolo}
  L&=&4\pi d_L^2 P_{\mathrm{bolo}}
  ,
  \\
  \label{eq:GRB-E_gamma-S_bolo}
  E_{\gamma}&=& E_{\gamma,\mathrm{iso}}F_{\mathrm{beam}}
  =4\pi d_L^2 S_{\mathrm{bolo}} (1+z)^{-1} F_{\mathrm{beam}}
  .
\end{eqnarray}
The conversion depends on cosmological parameters because
the luminosity distance $d_L$ depends on cosmological models. As a
result, the calibration and the constraining of cosmological
parameters are mixed for GRBs; i.e., we need to simultaneously fit
calibration parameters of GRBs and cosmological parameters.

Based on the above discussions, the $\chi^2$ value for GRBs is
calculated by
\begin{eqnarray}
  \label{eq:chi2_GRB}
  \chi_{\mathrm{GRB}}^2
  &=&
  \sum_i
  \frac{
    \left\{
      \log \frac{L_i}{1 \; \mathrm{erg} \; \mathrm{s}^{-1}}
      -a_1-b_1 \log
      \left[
        \frac{\tau_{\mathrm{lag},i}(1+z_i)^{-1}}{0.1\;\mathrm{s}}
      \right]
    \right\}^2
  }{\sigma_1^2}
  \nonumber \\
  && + \sum_i
  \frac{
    \left\{
      \log \frac{L_i}{1 \; \mathrm{erg} \; \mathrm{s}^{-1}}
      -a_2-b_2 \log
      \left[
        \frac{V_i(1+z_i)}{0.02}
      \right]
    \right\}^2
  }{\sigma_2^2}
  \nonumber \\
  && + \sum_i
  \frac{
    \left\{
      \log \frac{L_i}{1 \; \mathrm{erg} \; \mathrm{s}^{-1}}
      -a_3-b_3 \log
      \left[
        \frac{E_{\mathrm{peak},i}(1+z_i)}{300\;\mathrm{keV}}
      \right]
    \right\}^2
  }{\sigma_3^2}
  \nonumber \\
  && + \sum_i
  \frac{
    \left\{
      \log \frac{E_{\gamma,i}}{1\;\mathrm{erg}}
      -a_4-b_4 \log
      \left[
        \frac{E_{\mathrm{peak},i}(1+z_i)}{300\;\mathrm{keV}}
      \right]
    \right\}^2
  }{\sigma_4^2}
  \nonumber \\
  && + \sum_i
  \frac{
    \left\{
      \log \frac{L_i}{1 \; \mathrm{erg} \; \mathrm{s}^{-1}}
      -a_5-b_5 \log
      \left[
        \frac{\tau_{\mathrm{RT},i}(1+z_i)^{-1}}{0.1\;\mathrm{s}}
      \right]  
    \right\}^2
  }{\sigma_5^2}
  ,
\end{eqnarray}
where $L_i$ and $E_{\gamma,i}$ are derived using
Eq.~(\ref{eq:GRB-L-P_bolo}) and Eq.~(\ref{eq:GRB-E_gamma-S_bolo}). The
summations run over the GRBs with the corresponding luminosity
indicator observed.
We use the systematic errors estimated by~\citep{Schaefer:2006pa} that
account for the scatter of the log-log plots of the luminosity versus
the luminosity indicators as $\sigma_1$-$\sigma_5$ in our analysis.
Apparently, $\chi_{\mathrm{GRB}}^2$ is a function of calibration
parameters $a_1$-$a_5$, $b_1$-$b_5$ and cosmological parameters that
enter through the luminosity distance $d_L$. 

\subsection{Other data}
\label{sec:other_data}

In addition to SNe Ia and GRBs, we have also used the constraints
below
following previous analyses~\citep{Riess:2006fw,Sullivan:2007pd}
\begin{itemize}
\item \emph{Constraints on dimensionless mass densities}:
  The SDSS large-scale structure measurements give the constraint on
  local mass density in terms of
  $\Omega_m h =0.213 \pm 0.023$~\citep{Tegmark:2003uf}.
  The WMAP three-year data combined with the HST key project
  constraint on the Hubble constant gives
  $\Omega_k=-0.014 \pm 0.017$~\citep{Spergel:2006hy}.
\item \emph{The SDSS luminous red galaxy, baryon acoustic
    oscillation (BAO) distance parameter to $z_{\mathrm{BAO}}=0.35$}:
  $A \equiv \frac{\sqrt{\Omega_m H_0^2}}{c z_{\mathrm{BAO}}} 
  \left[ 
    r^2(z_{\mathrm{BAO}})
    \frac{c z_{\mathrm{BAO}}}{H_0 E(z_{\mathrm{BAO}})}
  \right]^{1/3}$,
  where $r(z)=d_L(z)/(1+z)$.
  $A=0.469\left(\frac{n}{0.98}\right)^{-0.35} \pm 0.017$
  from~\citet{Eisenstein:2005su} and the three-year WMAP results give
  $n=0.95$~\citep{Spergel:2006hy}.
\item \emph{The distance to last scattering, z=1089}:
  If nonzero cosmic curvature is allowed as we do in our analysis, the
  three-year WMAP data~\citep{Spergel:2006hy} gives the shift parameter
  $R_{\mathrm{CMB}} 
  = \frac{\sqrt{\Omega_m H_0^2}}{c} r(z_{\mathrm{CMB}})
  = 1.71 \pm 0.03$~\citep{Wang:2007mza}. 
\item \emph{The distance ratio between $z_{\mathrm{BAO}}=0.35$ and
    $z_{\mathrm{CMB}}=1089$}:
  \begin{equation}
    \label{eq:distance_ratio}
    R_{0.35}=\frac{
      \left[ 
        r^2(z_{\mathrm{BAO}})
        \frac{c z_{\mathrm{BAO}}}{H_0 E(z_{\mathrm{BAO}})}
      \right]^{1/3}
    }{r(z_{\mathrm{CMB}})}
    .
  \end{equation}
  The SDSS BAO analysis~\citep{Eisenstein:2005su} gives
  $R_{0.35}= 0.0979 \pm 0.0036$. 
\end{itemize}
The corresponding $\chi^2$ for these constraints are directly
calculated using Eq.~(\ref{eq:chi2_xi}).

We have also studied the dark energy EOS evolution with the above BAO
constraints replaced by the latest BAO measurements presented
in~\citet{Percival:2007yw},
for which the $\chi^2$ value is~\citep{Percival:2007yw}
\begin{equation}
  \label{eq:chi2_BAO}
  \chi_{\mathrm{BAO}}^2 =
  \boldsymbol{X}_{\mathrm{BAO}}^{\mathrm{T}}
  \boldsymbol{C}_{\mathrm{BAO}}^{-1}
  \boldsymbol{X}_{\mathrm{BAO}}
\end{equation}
where
\begin{equation}
  \label{eq:X_BAO}
  \boldsymbol{X}_{\mathrm{BAO}} =
  \left(
    \begin{array}{c}
      \frac{r_s}{D_V(0.2)} - 0.1980 \\
      \frac{r_s}{D_V(0.35)} - 0.1094
    \end{array}
  \right)
\end{equation}
with $r_s$ the comoving sound horizon at recombination and
\begin{equation}
  \label{eq:inv_cov_BAO}
  \boldsymbol{C}_{\mathrm{BAO}}^{-1} =
  \left(
    \begin{array}{cc}
      35059   &  -24031 \\
      -24031  &  108300
    \end{array}
  \right).
\end{equation}
This constraint itself favors a dark energy EOS of
$w<-1$~\citep{Percival:2007yw}.

\section{Results}
\label{sec:results}

\begin{figure}[htbp]
  \centering
  \includegraphics[width=0.37\textwidth]{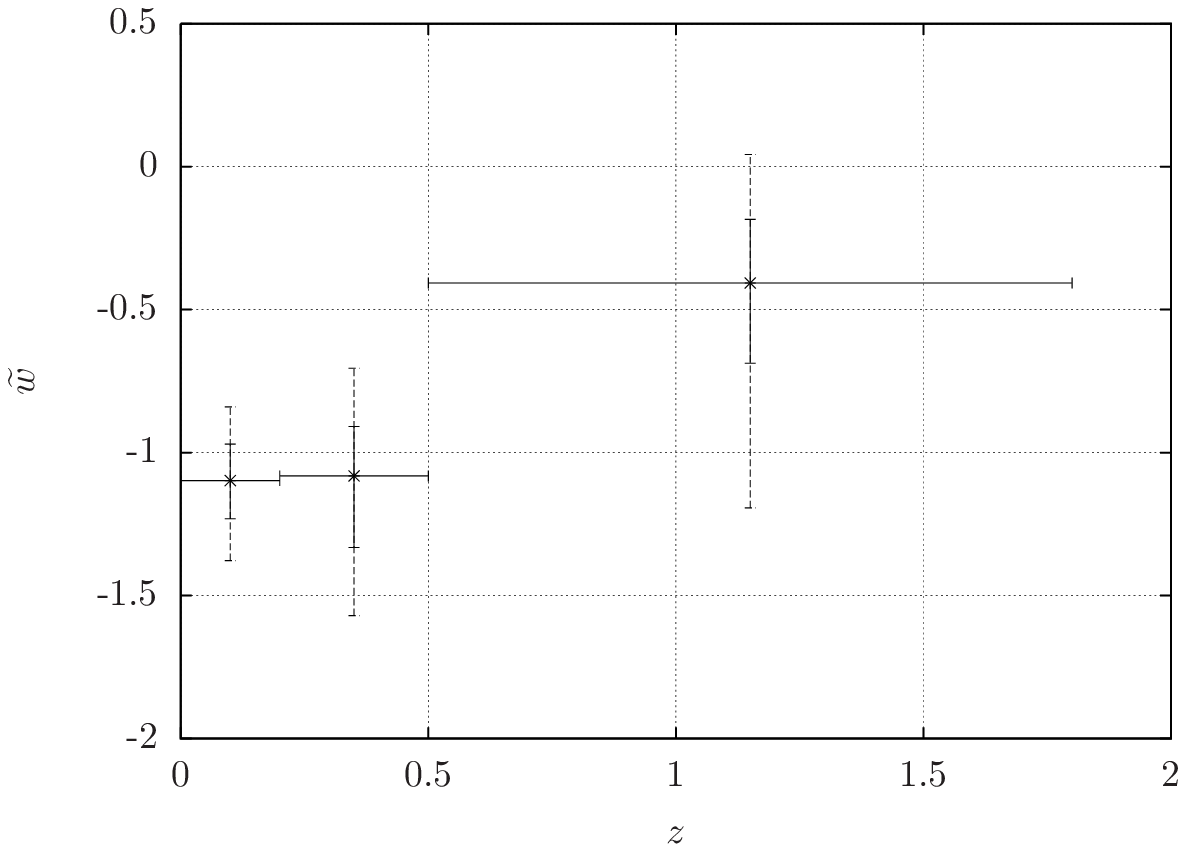}
  \includegraphics[width=0.37\textwidth]{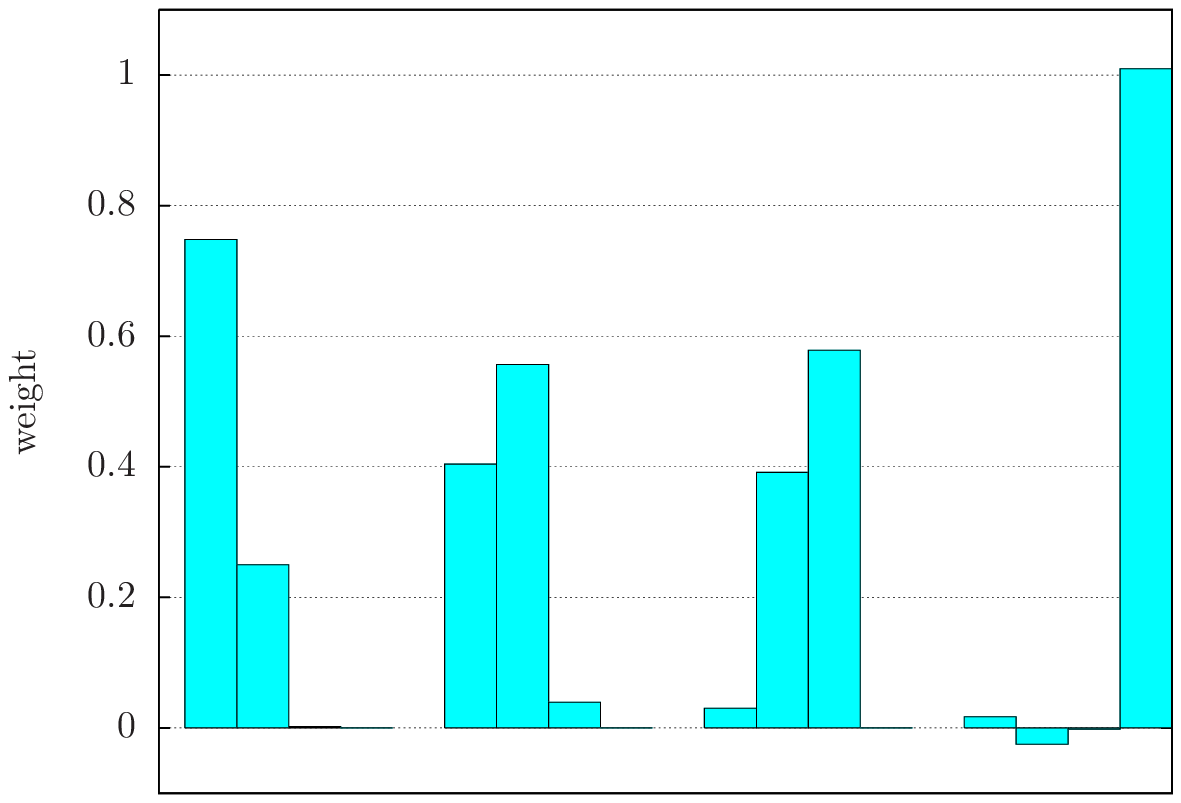}
  \includegraphics[width=0.37\textwidth]{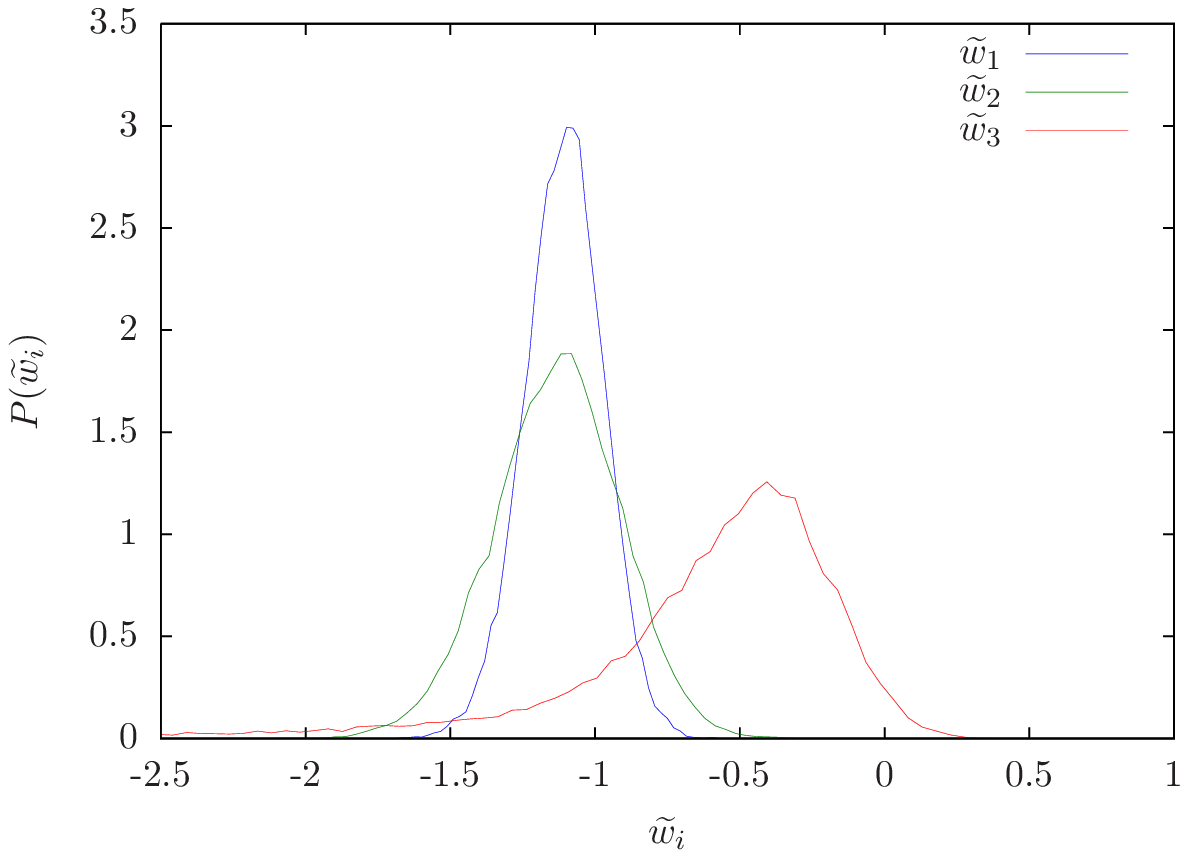}
  \includegraphics[width=0.37\textwidth]{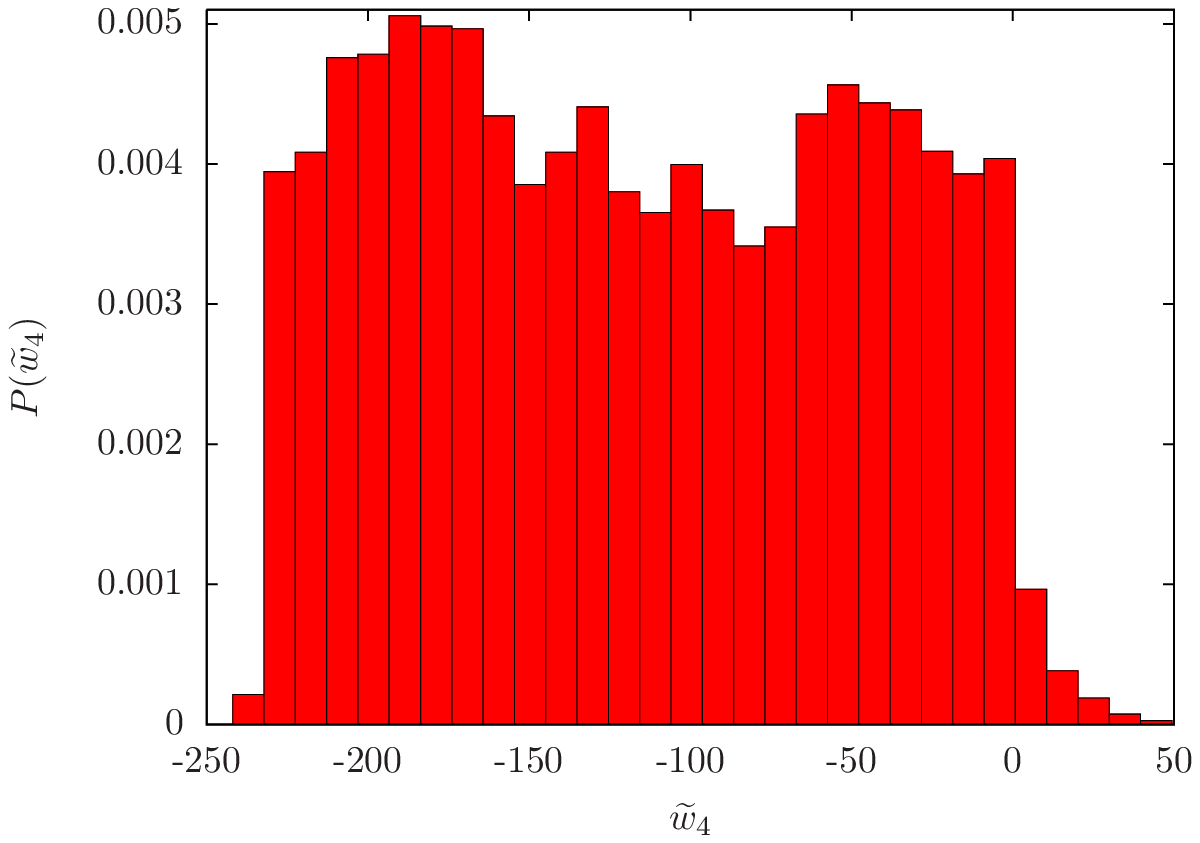}
  \includegraphics[width=0.37\textwidth]{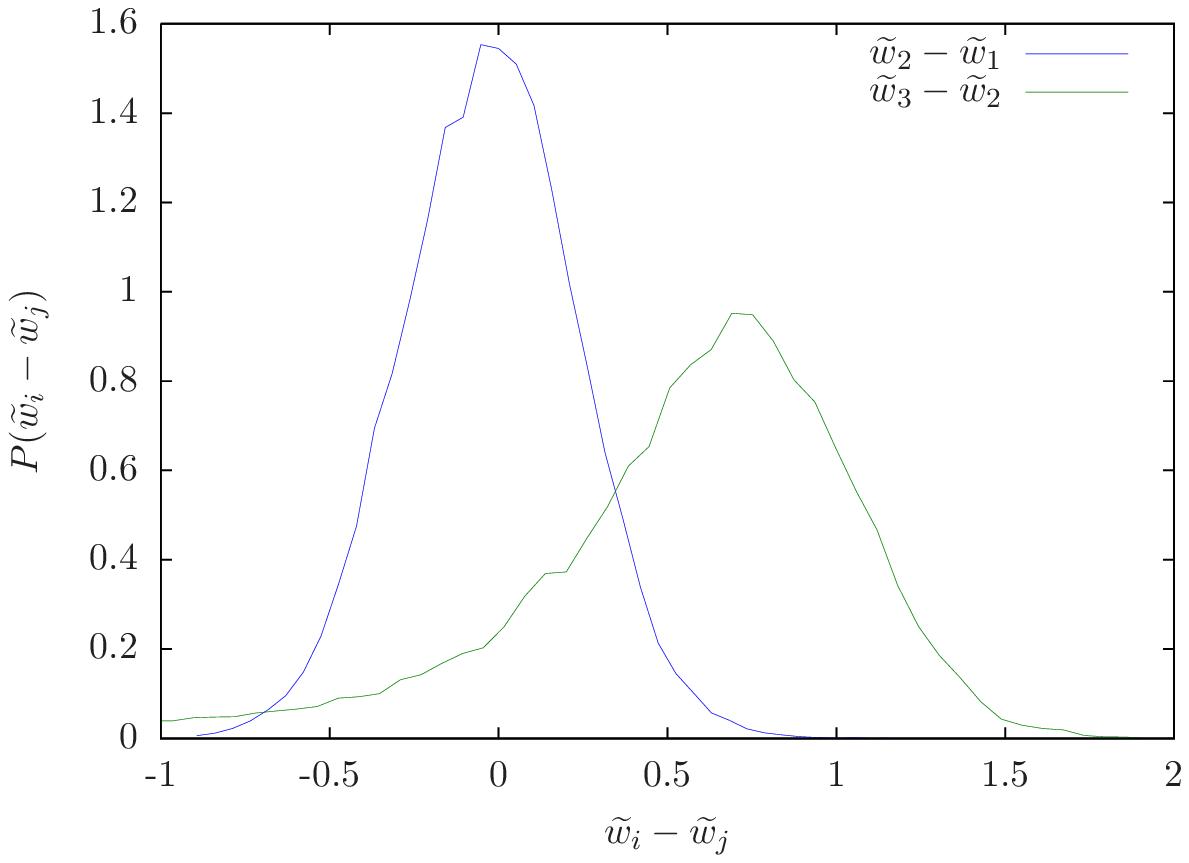}
  \caption{Estimates of the uncorrelated dark energy EOS parameters
    using the weak prior. In turn are the plots of $\widetilde{w}_i$
    ($i=1$-$3$) versus redshift, window functions of $\widetilde{w}_i$
    ($i=1$-$4$) with respect to the $4$ bins, probability distribution
    of $\widetilde{w}_i$ ($i=1$-$3$), $\widetilde{w}_4$, and
    $\widetilde{w}_i-\widetilde{w}_j$.}
  \label{fig:weak_prior_result}
\end{figure}

\begin{figure}[htbp]
  \centering
  \includegraphics[width=0.37\textwidth]{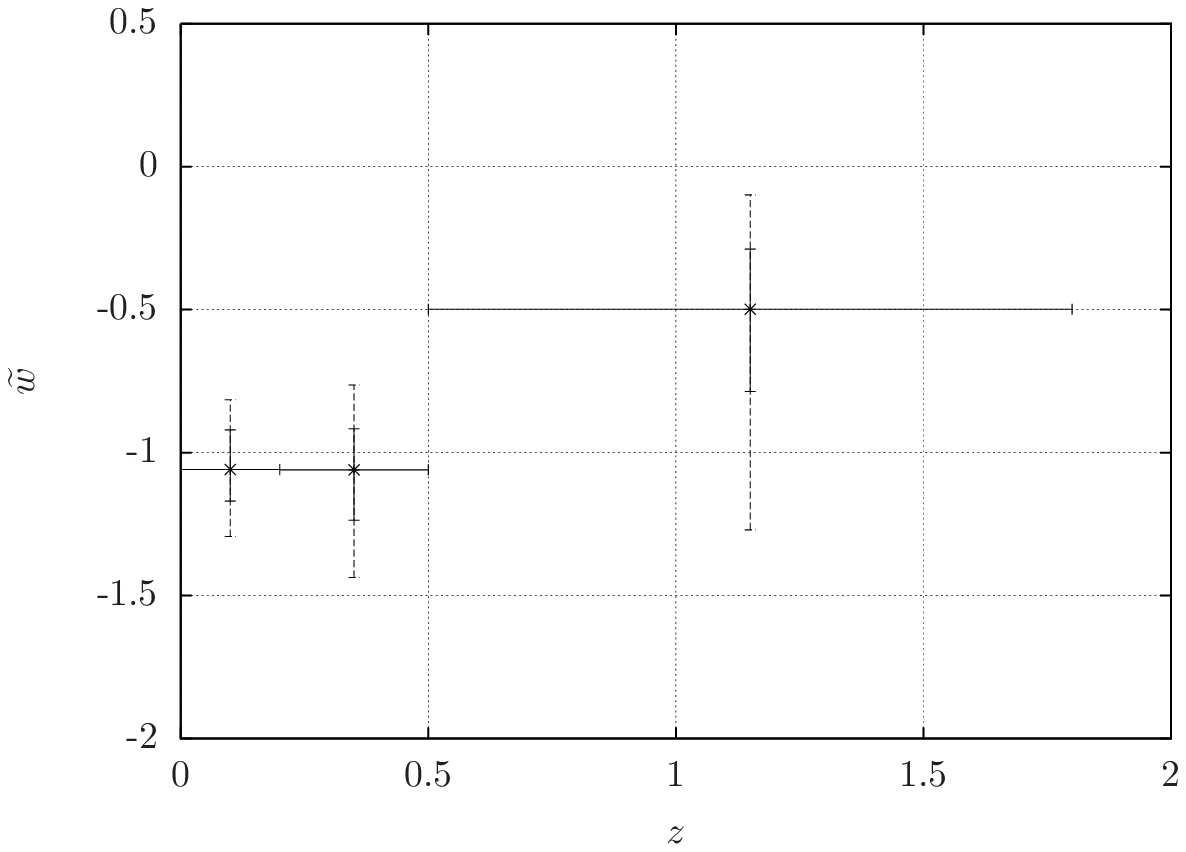}
  \includegraphics[width=0.37\textwidth]{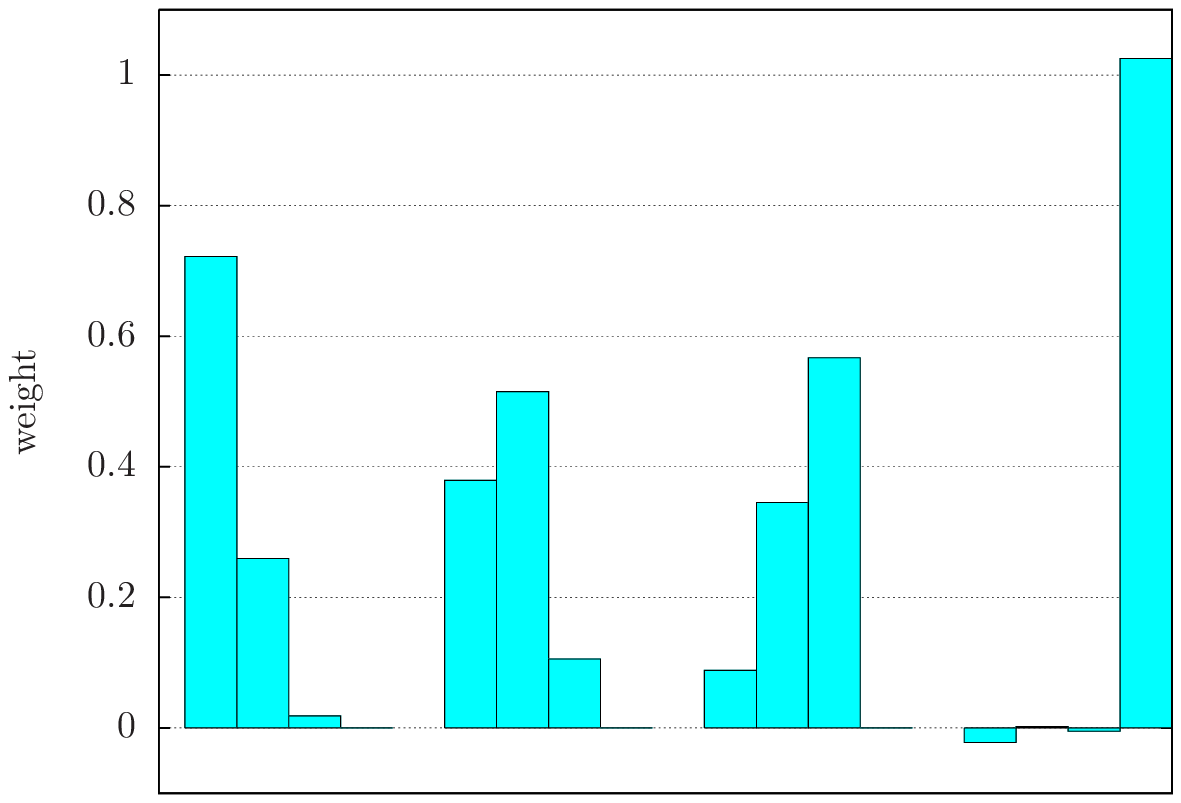}
  \includegraphics[width=0.37\textwidth]{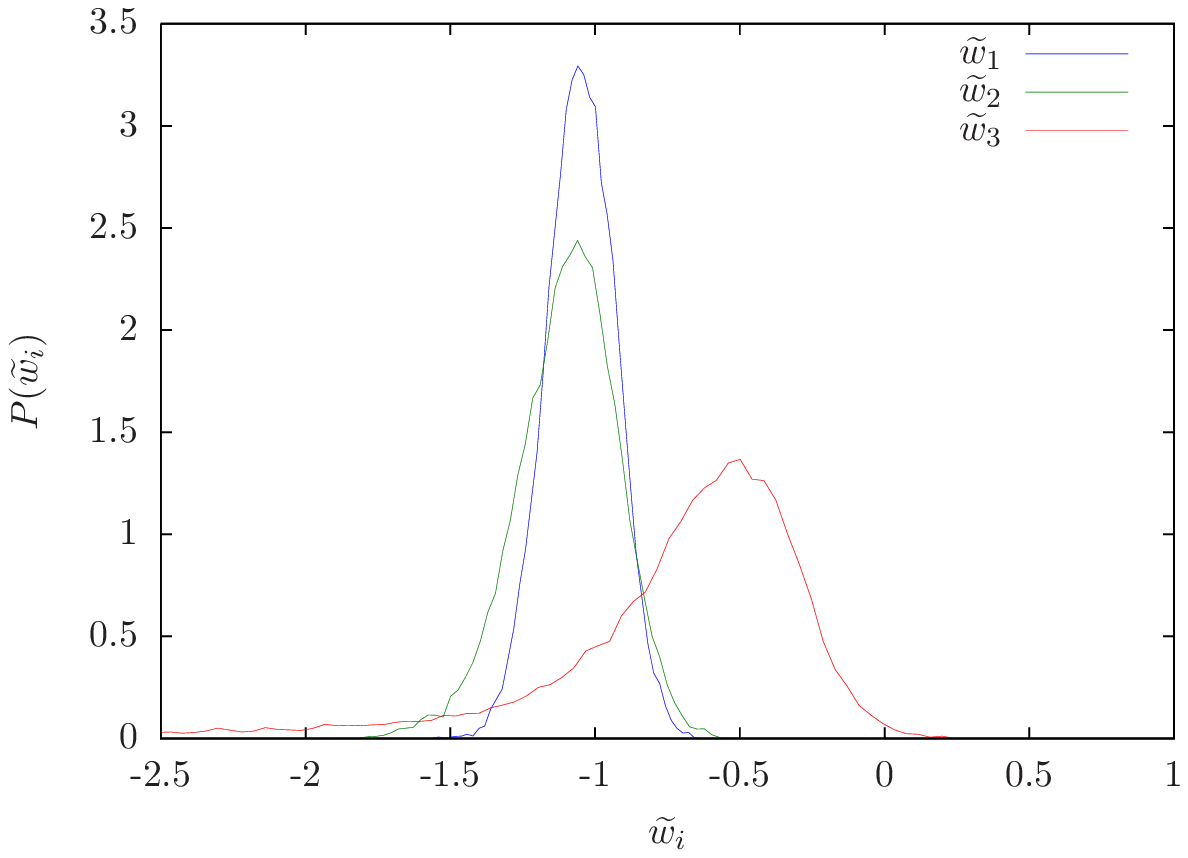}
  \includegraphics[width=0.37\textwidth]{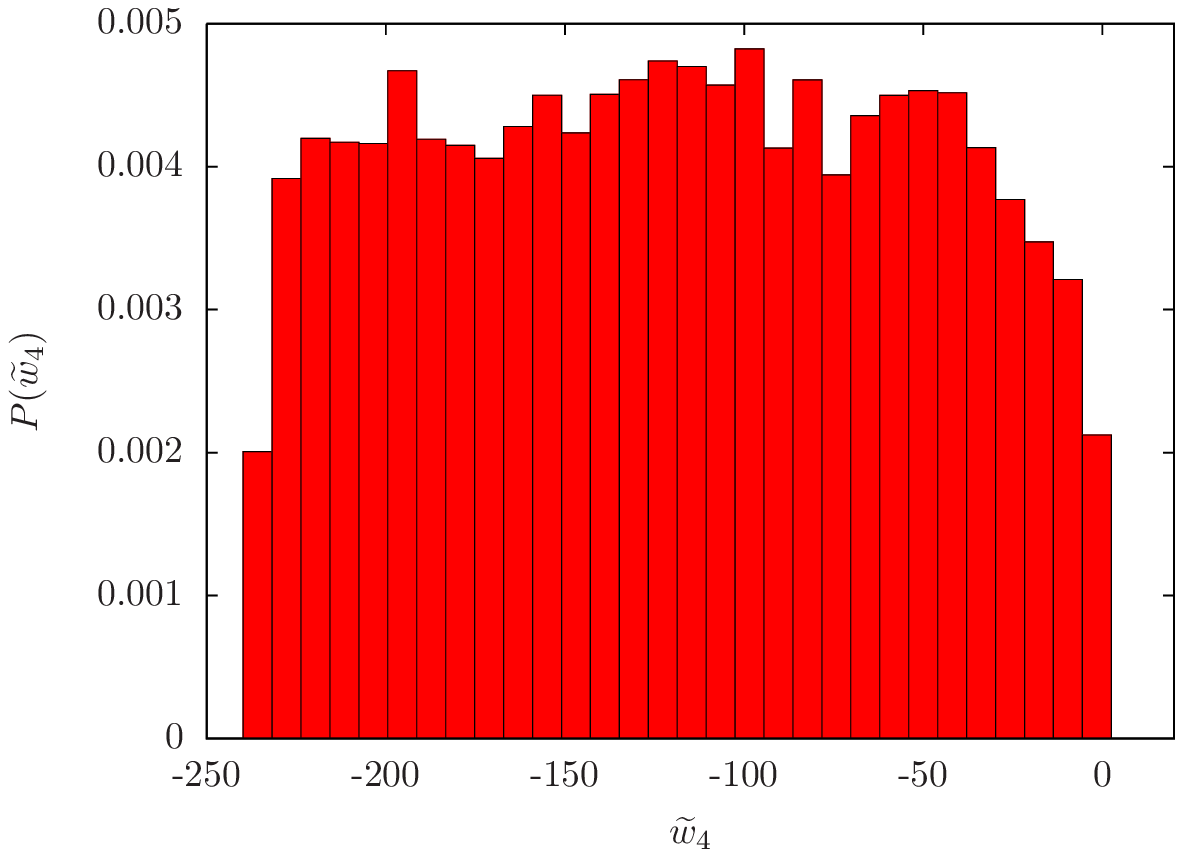}
  \includegraphics[width=0.37\textwidth]{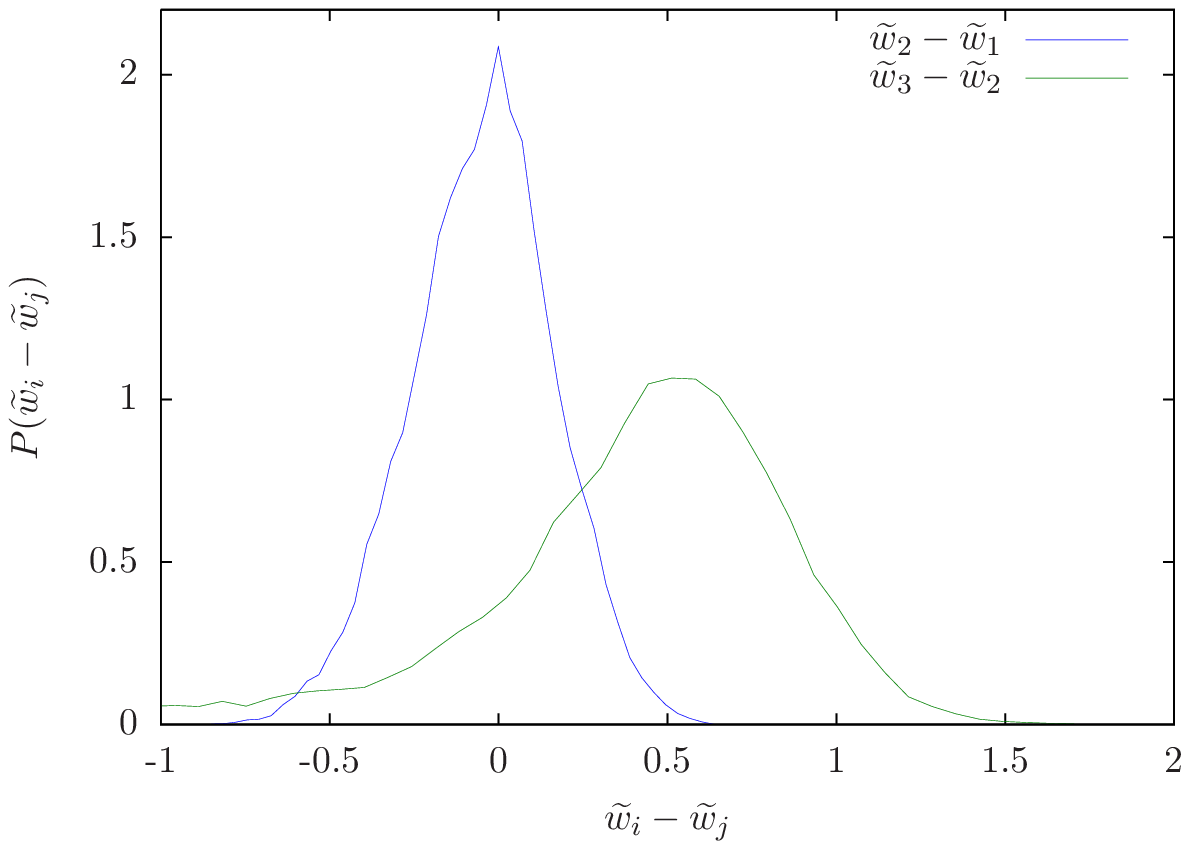}
  \caption{Estimates of the uncorrelated dark energy EOS parameters
    using the strong prior. Same as Figure~\ref{fig:weak_prior_result}
    except using the strong prior.}
  \label{fig:strong_prior_result}
\end{figure}

\begin{figure}[htbp]
  \centering
  \includegraphics[width=0.37\textwidth]{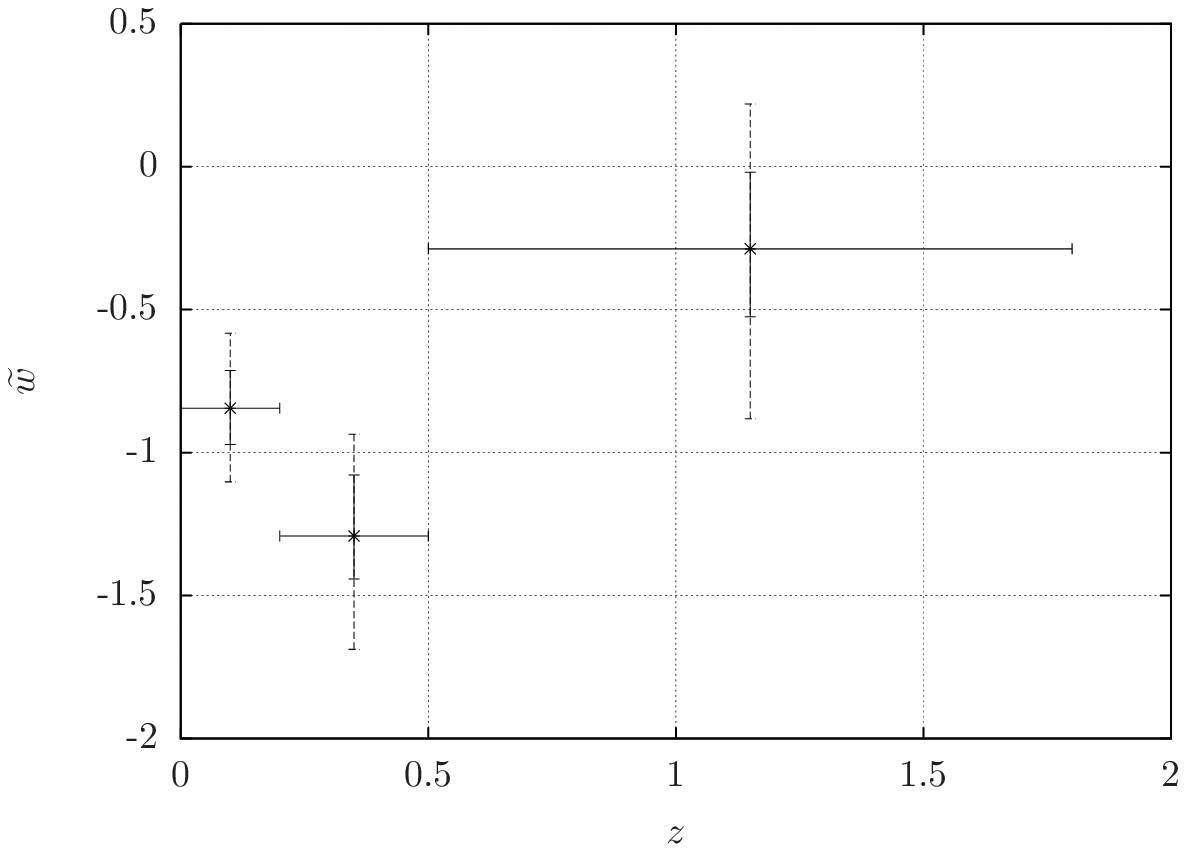}
  \includegraphics[width=0.37\textwidth]{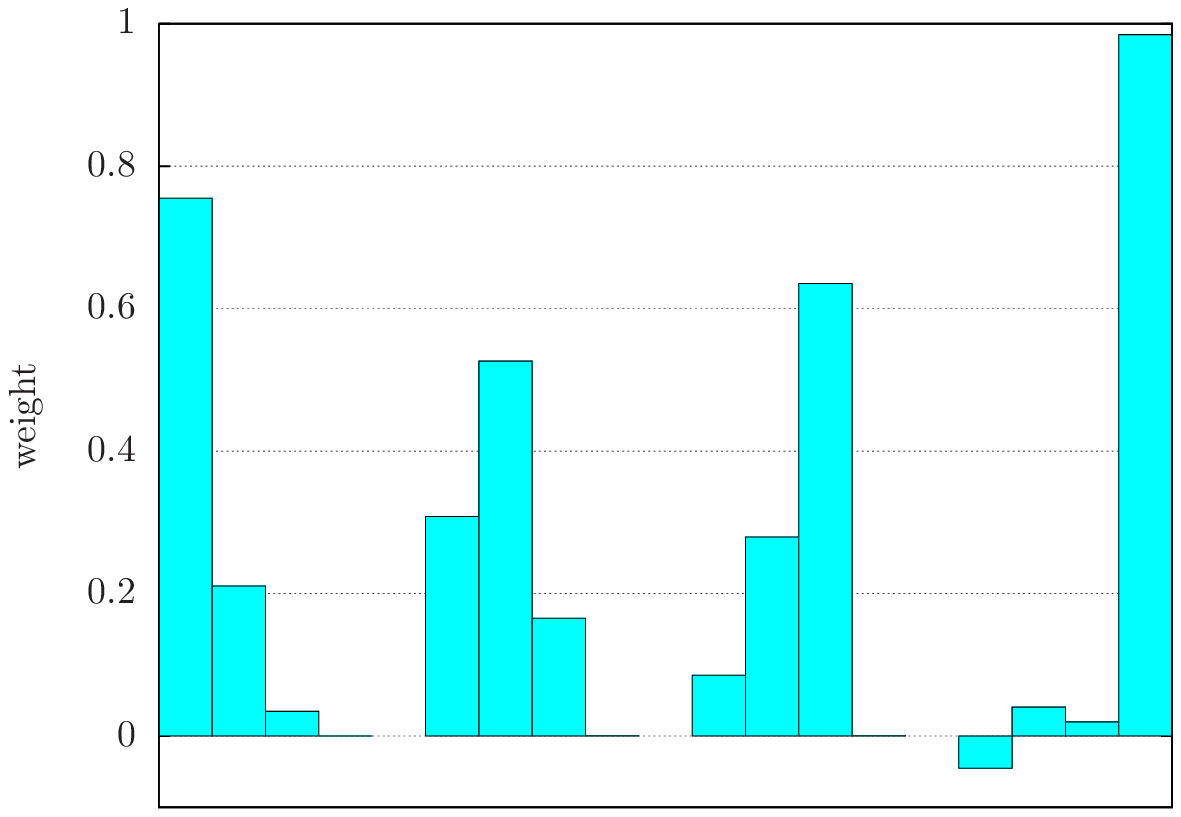}
  \includegraphics[width=0.37\textwidth]{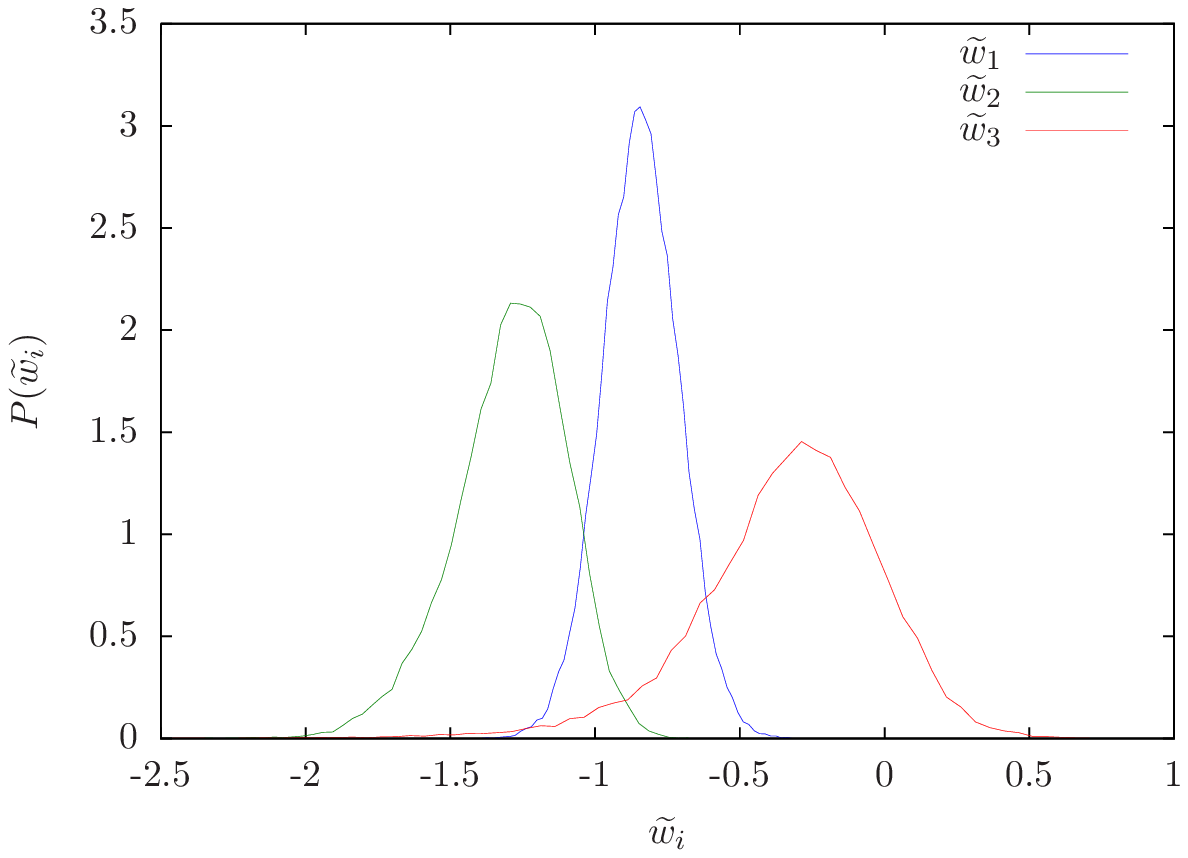}
  \includegraphics[width=0.37\textwidth]{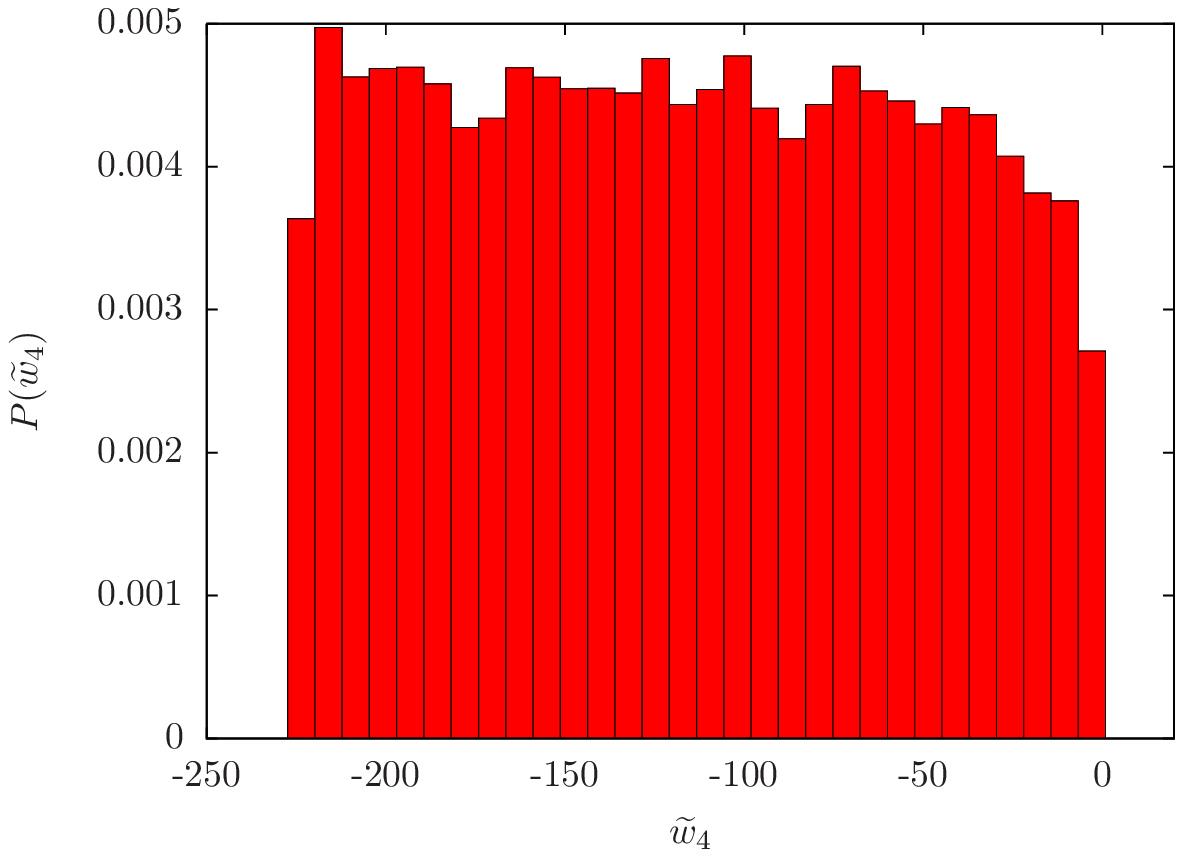}
  \includegraphics[width=0.37\textwidth]{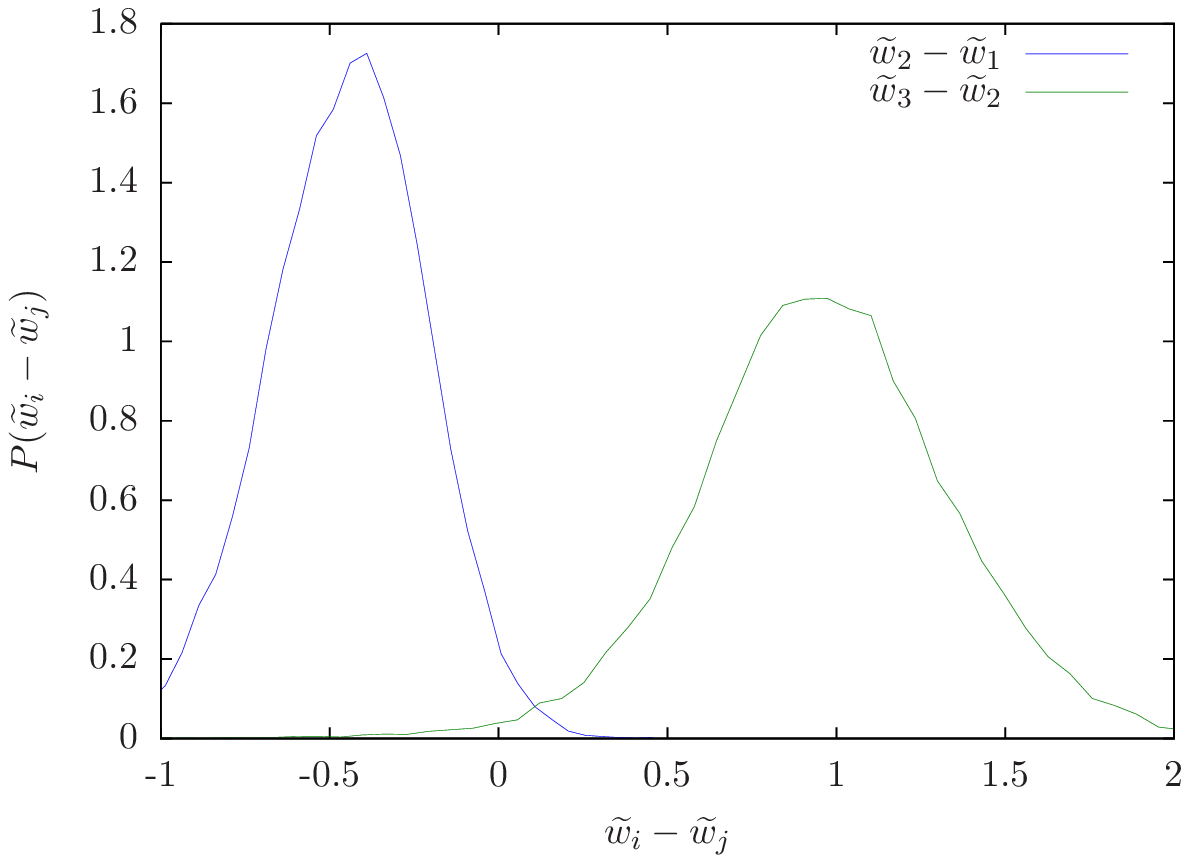}
  \caption{Estimates of the uncorrelated dark energy EOS parameters
    using the strong prior. Same as Figure~\ref{fig:strong_prior_result}
    except BAO constraints were updated with the latest
    measurements~\citep{Percival:2007yw}.}
  \label{fig:new_BAO_result}
\end{figure}

Figures~\ref{fig:weak_prior_result} and~\ref{fig:strong_prior_result}
show our results for the weak prior and strong prior
respectively. For these two figures, we have included subsets of data
from section~\ref{sec:other_data} same as that are used
in~\citet{Sullivan:2007pd} besides SNe Ia. For the results presented
in Figure~\ref{fig:new_BAO_result}, the BAO constraints are updated
with the latest measurements~\citep{Percival:2007yw}, see
Eq.~(\ref{eq:chi2_BAO}), Eq.~(\ref{eq:X_BAO}), and
Eq.~(\ref{eq:inv_cov_BAO}).
A comparison between Figures~\ref{fig:weak_prior_result}
and~\ref{fig:strong_prior_result} shows that the results
are insensitive to the priors, i.e. insensitive to whether $w(z>7)=-1$
is assumed or not for dark energy.

Since Figures~\ref{fig:weak_prior_result}
and~\ref{fig:strong_prior_result} only differ from results derived 
by~\citet{Sullivan:2007pd} in that we include GRB luminosity data,
comparisons of Figures~\ref{fig:weak_prior_result}
and~\ref{fig:strong_prior_result} with figures
in~\citet{Sullivan:2007pd} demonstrate the improvement made by
including GRBs. We find
that there is little improvement in $\widetilde{w}_1$ and
$\widetilde{w}_2$. This is because at low redshift, where we have
both SNe Ia and GRBs, there are fewer GRBs than that of
SNe Ia (see Table~\ref{tab:distr_sg}, in the first two bins the number
of GRBs is negligible compared with that of SNe Ia); at the same time,
the contributions to $\widetilde{w}_1$ and $\widetilde{w}_2$ from high
redshift, where we have a considerable number of GRB samples (see
Table~\ref{tab:distr_sg}), are too small
(see the weight histograms in Figures~\ref{fig:weak_prior_result}
and~\ref{fig:strong_prior_result})
to improve constraints on $\widetilde{w}_1$ and $\widetilde{w}_2$
significantly.
\begin{table}[htbp]
  \centering
  \begin{tabular}{|c|c|c|c|c|}
    \hline
    bin & 1 & 2 & 3 & 4 \\
    \hline \hline
    redshift range & 0-0.2 & 0.2-0.5 & 0.5-1.8 & 1.8-7 \\
    \hline
    number of SNe Ia & 47 & 59 &  86 &  0 \\
    \hline
    number of GRBs   &  1 &  3 &  32 & 33 \\
    \hline
    total number     & 48 & 62 & 118 & 33 \\
    \hline
  \end{tabular}
  \caption{Number of SNe Ia and GRBs that fall into the four bins}
  \label{tab:distr_sg}
\end{table}
The most significant improvement lies in $\widetilde{w}_3$, whose
contribution mostly comes from the third bin, where we have
several GRBs (see Table~\ref{tab:distr_sg}). The
$1\sigma$ confidence interval of $\widetilde{w}_3$ with GRBs included
is less than one third of that presented in~\citet{Sullivan:2007pd}
without including GRB luminosity data.

For Figures~\ref{fig:weak_prior_result}
and~\ref{fig:strong_prior_result}, $\widetilde{w}_1$ and
$\widetilde{w}_2$ are consistent with the cosmological constant within
$1\sigma$, and $\widetilde{w}_3$ consistent within $2\sigma$. While
in Figure~\ref{fig:new_BAO_result}, for which the latest BAO
measurements are used instead, the
cosmological constant lies outside of the $1\sigma$ confidence
intervals of $\widetilde{w}_1$ and $\widetilde{w}_2$, and outside the
$2\sigma$ confidence interval of $\widetilde{w}_3$, though still
inside the $2\sigma$ confidence intervals of $\widetilde{w}_1$ and
$\widetilde{w}_2$. These results show
some evidence of an evolving dark energy EOS.
This is not surprising provided that the latest BAO measurements
themselves favor a dark energy EOS of $w<-1$~\citep{Percival:2007yw}.
The BAO distance information lies in the second redshift bin, so
including it leads to a smaller $\widetilde{w}_2$.
And main data we used
depends on the integration of the dark energy evolution, thus the
decrease in $\widetilde{w}_2$ causes increases in $\widetilde{w}_1$
and $\widetilde{w}_3$.

The constraints on $\widetilde{w}_4$ are very weak. The uncertainty is
so great that we plot its probability separately. This is due to
three reasons. First, there are not enough samples of standard
candles in the fourth bin, all of which are GRBs.
From Table~\ref{tab:distr_sg} it can be seen that the number
ratio of third bin to the fourth bin is about $4$. Second, as
mentioned earlier, the estimate of the behavior of dark energy at
high redshift depends on its behavior at low redshift; consequently,
the uncertainty of EOS parameters at low redshift
will be reflected on EOS parameters at high redshift. Therefore we
get increasing errors as the redshift increases. Thirdly, the density
ratio of dark energy to matter is given by (assuming a constant EOS
parameter for dark energy)
\begin{equation}
  \label{eq:ratio_de_dm}
  \frac{\rho_x}{\rho_m}
  =\frac{\rho_{x0}(1+z)^{3(1+w_x)}}{\rho_{m0}(1+z)^3}
  \approx 3(1+z)^{3w_x}
  .
\end{equation}
For negative $w_x$, the ratio decreases as $z$ increases. For example,
when $w_x=-1$, the ratio is about $1/9$ at $z=2$. At higher redshift,
matter dominates over dark energy, then dark energy becomes less
important in determining the cosmic expansion. Thus the constraints
imposed on the behavior of dark energy by the expansion history become
weak compared with that at low redshift where dark energy is
important. Despite the large uncertainty in $\widetilde{w}_4$, there
is indeed some restriction imposed by GRBs. From the
probability plots of $\widetilde{w}_4$ in
Figures~\ref{fig:weak_prior_result}, \ref{fig:strong_prior_result},
and~\ref{fig:new_BAO_result}, it can be seen that there is 
obviously a cut at about zero. In other words, it is most probable
that the ratio in Eq.~(\ref{eq:ratio_de_dm}) 
continues to decrease at a redshift beyond $1.8$.
The probability cut at the left of $-200$ is due to the precision of
the computer and can be viewed as the negative infinity.
To get substantial constraints on the dark energy EOS beyond $1.8$, we
need more GRB samples.

To see the overall improvement made by including GRB luminosity data,
we calculate the figure of merit (FOM), which is defined
by~\citep{Sullivan:2007pd, Sullivan:2007tx}
\begin{equation}
  \label{eq:fom}
  \mathrm{FOM} =
  \left[
    \sum_i \frac{1}{\sigma^2(\widetilde{w}_i)}
  \right]^{1/2}
  .
\end{equation}
For the the results presented in Figure~\ref{fig:new_BAO_result},
$\mathrm{FOM} = 9.6$. And if the GRB luminosity data are excluded,
$\mathrm{FOM} = 8.8$.

\section{Summary}
\label{sec:summary}

We used a model-independent approach to constrain the evolution of
dark energy. First, we separated the redshifts into $4$ bins and
assumed
a constant EOS parameter for dark energy in each bin, then estimated
the
uncorrelated EOS parameters. We mainly used the SNe Ia and GRBs in our
analysis. Other constraints from SDSS, 2dFGRS, HST, and WMAP are also
included. Compared with the results obtained without including GRB
luminosity data, the confidence
interval of the third uncorrelated EOS parameter, whose
contribution mostly comes from the third bin, is reduced
significantly. Even though constraints at high redshift where we have
only GRBs are very weak, from the obvious probability cut of the EOS
parameter at about zero, we can infer that it is most probable that
the ratio of dark energy to
matter continues to decrease beyond redshift $1.8$. 
To get substantial constraints at redshifts beyond SNe Ia more GRBs
are needed.

If the latest BAO measurements, which themselves favor a dark energy
EOS of $w<-1$, are included, the results show some evidence for an
evolving dark energy EOS. Otherwise, the results are consistent with
the cosmological constant.

\begin{acknowledgements}
Shi Qi would like to thank Maurice HPM van Putten and Edna Cheung for
helpful discussions and suggestions.
This work was supported by the Scientific Research Foundation of the
Graduate School of Nanjing University (for Shi Qi),
the Jiangsu Project Innovation
for PhD Candidates CX07B-039z (for Fa-Yin Wang),
and the National Natural Science Foundation of China
under Grant No.~10473023.
\end{acknowledgements}

\bibliographystyle{aa}
\bibliography{dark_energy}

\end{document}